\documentclass[prb,twocolumn,floatfix,showpacs,superscriptaddress]{revtex4} 
\usepackage{graphicx} 
\usepackage{amsmath} 
\begin{document}  
\title{Classical trajectories in quantum transport at the band center of bipartite lattices with or without vacancies} 
\author{G. Chiappe} 
\affiliation{Departamento de F\'{\i}sica Aplicada and Unidad Asociada del Consejo Superior de Investigaciones Cient\'{\i}ficas, Universidad de Alicante, San Vicente del Raspeig, Alicante 03690, Spain.}  
\affiliation{Departamento de F\'{\i}sica  J.J. Giambiagi, Facultad de Ciencias Exactas y Naturales, Universidad de Buenos Aires, Ciudad Universitaria, 1428 Buenos Aires, Argentina} 
\author{E. Louis} 
\affiliation{Departamento de F\'{\i}sica Aplicada and Unidad Asociada del Consejo Superior de Investigaciones Cient\'{\i}ficas, Universidad de Alicante, San Vicente del Raspeig, Alicante 03690, Spain.}
\author{M.J. S\'anchez}
\affiliation{Departamento de F\'{\i}sica  J.J. Giambiagi, Facultad de Ciencias Exactas y Naturales, Universidad de Buenos Aires, Ciudad Universitaria, 1428 Buenos Aires, Argentina} 
\author{J. A. Verg\'es} \affiliation{Departamento de Teor\'{\i}a de la Materia Condensada, Instituto de Ciencia de Materiales de Madrid (CSIC), Cantoblanco, Madrid 28049, Spain.} 
\date{\today} 
\begin{abstract} 
Here we report on several anomalies in quantum transport at the band 
center of a bipartite lattice with vacancies that are surely due
to its chiral symmetry, namely: no weak localization effect shows up,
and, when leads have a single channel the transmission is
either one or zero.
We propose that these are a consequence of both the chiral symmetry and the 
large number of states at the band center. The probability  amplitude
associated to the  eigenstate that gives unit transmission ressembles a classical trajectory both with or without vacancies. The  large number of states
allows to build up trajectories that
elude the blocking vacancies explaining the absence of  weak localization.
\end{abstract} 
\maketitle  

\noindent {\it Bipartite lattices.} The possibility that qualitative differences between models with pure diagonal or non-diagonal disorder may exist, have attracted a considerable interest since Dyson's work on  a phonon model in one dimension \cite{Dy53}. For instance it has been reported that  non-diagonal disorder promotes a delocalization transition at $E=0$ \cite{BM98,Ve01,GM98}. 
Although this raised a  controversy regarding the general statement saying that the particular type of disorder should not matter in a single parameter scaling theory \cite{Ec83}, in recent years a different  view has emerged which ascribes  that transition to
the peculiar properties of bipartite lattices. These lattices are characterised by the electron-hole symmetry of the spectrum  (also known as chiral symmetry) and, while  diagonal disorder efficiently breaks this symmetry, pure non-diagonal disorder does not. More recently, however, it has been shown \cite{CS03}
that, at $E=0$, standard exponential 
localization occurs in a system with vacancies (a defect that 
also preserves chirality), reopening the mentioned controversy. 

Here we report on  some  anomalies in transport
properties at the band center of bipartite lattices that
are indeed related to chirality, namely: 
i) the absence of weak localization \cite{MB00,LV01}, and, ii) when single 
channels leads are connected 
to  cavities with or without vacancies, the transmission is 
either zero or one.  We show that these anomalies are a associated to the  
existence of  ballistic classical trajectories which
are possible due to  chiral symmetry and the large number of states at  $E=0$.

We illustrate these ideas on cavities of the square lattice  with vacancies.
\begin{figure} 
\includegraphics[width=2.5in,height=3in]{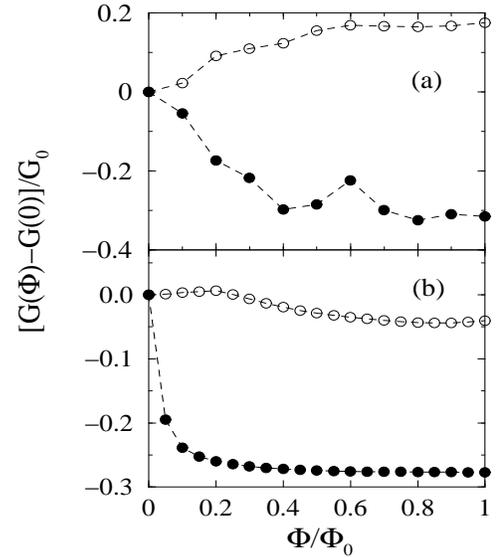} 
\caption{Change in the conductance versus magnetic flux (both in units of their respective quanta) in: a) $78 \times 78$ clusters with 78 vacancies (1200 realizations were included) and leads of width  $W=4$ connected at opposite corners of the cavity, and b) $6 \times 6$ clusters with 6 vacancies  (all realizations were included in the calculation) and leads of width 1 connected at opposite corners of the cavity. The results correspond to $E=0$ (filled circles) and $E=-0.1$ (empty circles). \label{fi_78_6}} 
\end{figure}  
\noindent {\it The Hamiltonian.} We consider a tight-binding Hamiltonian in $L \times L$ clusters of the square lattice with a single atomic orbital  per lattice site, \begin{eqnarray} {\hat H} = \sum_{m,n} \epsilon_{m,n} |m,n><m,n|- \nonumber \\ \sum_{<mn;m'n'>}t_{m,n;m'n'}|m,n><m',n'| \end{eqnarray} \noindent where $|m,n>$ represents an atomic orbital at site $(m,n)$,
and $\epsilon_{m,n}$ its energy (in the perfect system all were
taken equal to zero). For zero magnetic field the hopping energies between
nearest-neighbor sites (the symbol $<>$ denotes that the sum is restricted
to nearest-neighbors) were taken equal to 1.
A vacancy was introduced at site $(m,n)$  by taking $\epsilon_{m,n}=\infty$.
On the other hand, for finite magnetic field, and using Landau's gauge, the hopping energy from site $(m,n)$ to site (m',n') is written as $t_{m,n;m'n'}={\rm exp}\{2\pi i[m/(L-1)^2](\Phi/\Phi_0)\}$, where $\Phi$ is the magnetic flux and $\Phi_0$ the magnetic flux quantum. In the absence of defects this system has  $L$ eigenstates at the band center with momentum ${\bf k}=(k_x,k_y)$ such that $k_x+k_y=\pi$ (the lattice constant  is hereafter taken
as the length unit).  Once the cavity was connected to leads of width $W$,
the conductance $G$ was calculated by means 
of Green's function techniques \cite{Ve99}.
\begin{figure} 
\includegraphics[width=2.5in,height=3in]{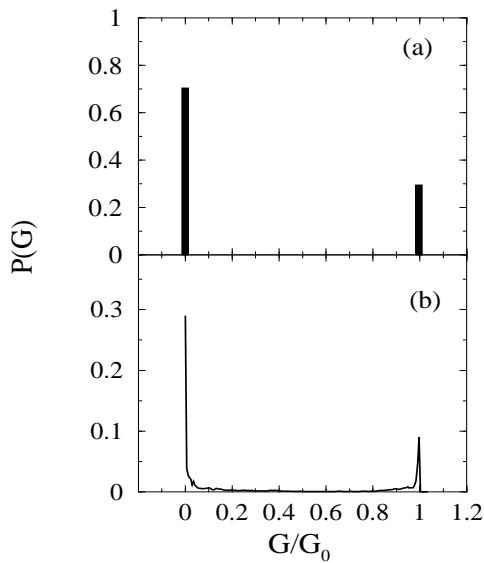} 
\caption{Conductance distribution in  $6 \times 6$ cavities with  6 vacancies (all realizations were included) and leads of width 1 connected at opposite corners of the cavity. The results correspond to $E=0$ (a) and $E=-0.1$ (b). \label{pd_6}} 
\end{figure}  
\noindent {\it Quantum transport anomalies in bipartite lattices  with vacancies.} In Fig. \ref{fi_78_6} we plot the results for the conductance versus magnetic flux in two cases.   In the upper panel results for a rather large cavity with leads of width $W=4$ are reported. It can be noted that while at an energy close to the band center the weak localization effect clearly shows up,  just at the band center the conductance unambigously decreases with the magnetic flux \cite{LV01}. The lower panel of the figure provides 
further support to this result. It shows the results of a calculation on a rather small cluster which includes all possible configurations of disorder. Again at the band center no weak localization effect is present. At $E=-0.1$, instead, we observe a clean increase of $G$ up to a flux in which an abrupt change of slope is noted. The latter is probably an effect of level crossing.  In any case, this results confirms that obtained on the larger cluster.
The conductance distribution obtained on $6 \times 6$ clusters with 6 vacancies and  leads of width 1 connected at opposite corners of the cavity are shown in Fig. \ref{pd_6}. All disorder realizations were included in the calculation. At the band center the distribution is reduced to two $\delta$-functions at $G$=0 and $G_0$, {\it i.e.}, the cavity  either does not conduct or behaves as a purely  ballistic system with unit transmission.  Similar behavior is obtained 
for other cluster sizes and averaging on different disorder configurations.
This is no longer the case at a finite energy (still quite close to the band center) as shown in the lower panel of the figure.  Fig. \ref{pd_6_30} illustrates how increasing the leads width and  cavity size affect the rather odd result found at $E=0$ \cite{CS03}.  The conductance distribution for the $6 \times 6$ cavity with $W=6$ leads and 6 vacancies, 
although still show peaks at $G=G_0, 2G_0, ..$ they are no longer  $\delta$-functions. Increasing the system size shows the expected  tendency towards a gaussian distribution. The probable causes of these results are discussed below.  \\ 
\begin{figure}
\includegraphics[width=2.8in,height=3in]{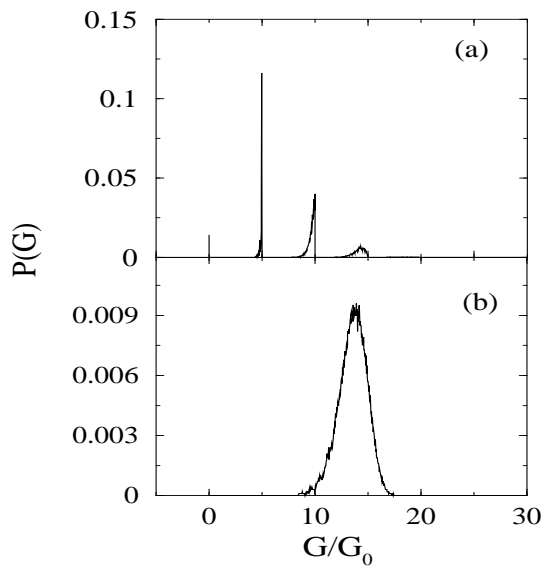}
\caption{Conductance distribution in 
$L \times L$ cavities with $L$ vacancies connected to leads of width L.
The results correspond to $L=6$ (a) and $L=30$ (b). For $L=6$ all
realizations were included while for $L=30$ the distribution
was calculated with 300000 realizations. The $x$-axis in the upper panel
should be divided by 5.
\label{pd_6_30}}
\end{figure}

\begin{figure}
\includegraphics[width=4in,height=3in]{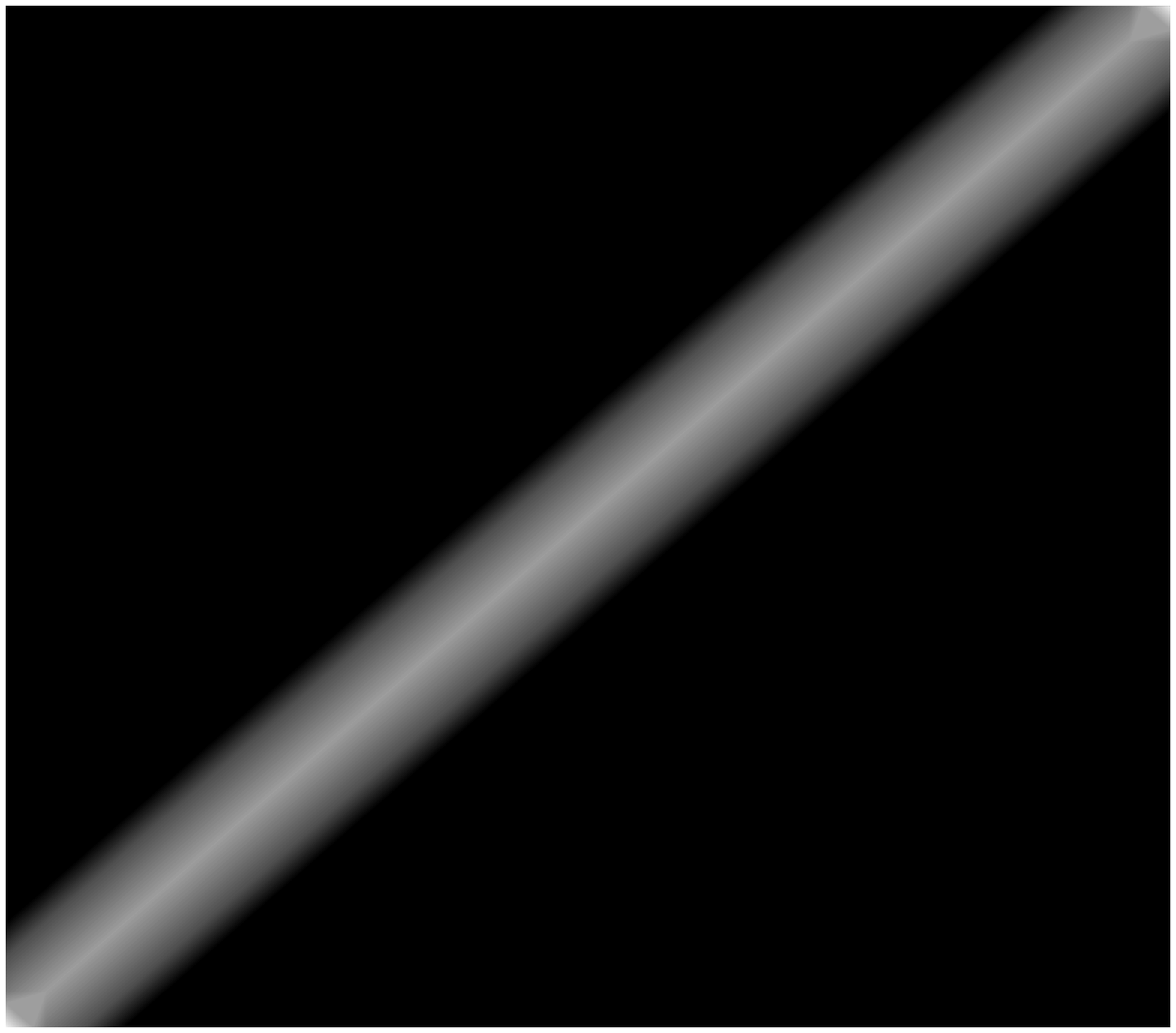} 
\includegraphics[width=4in,height=3in]{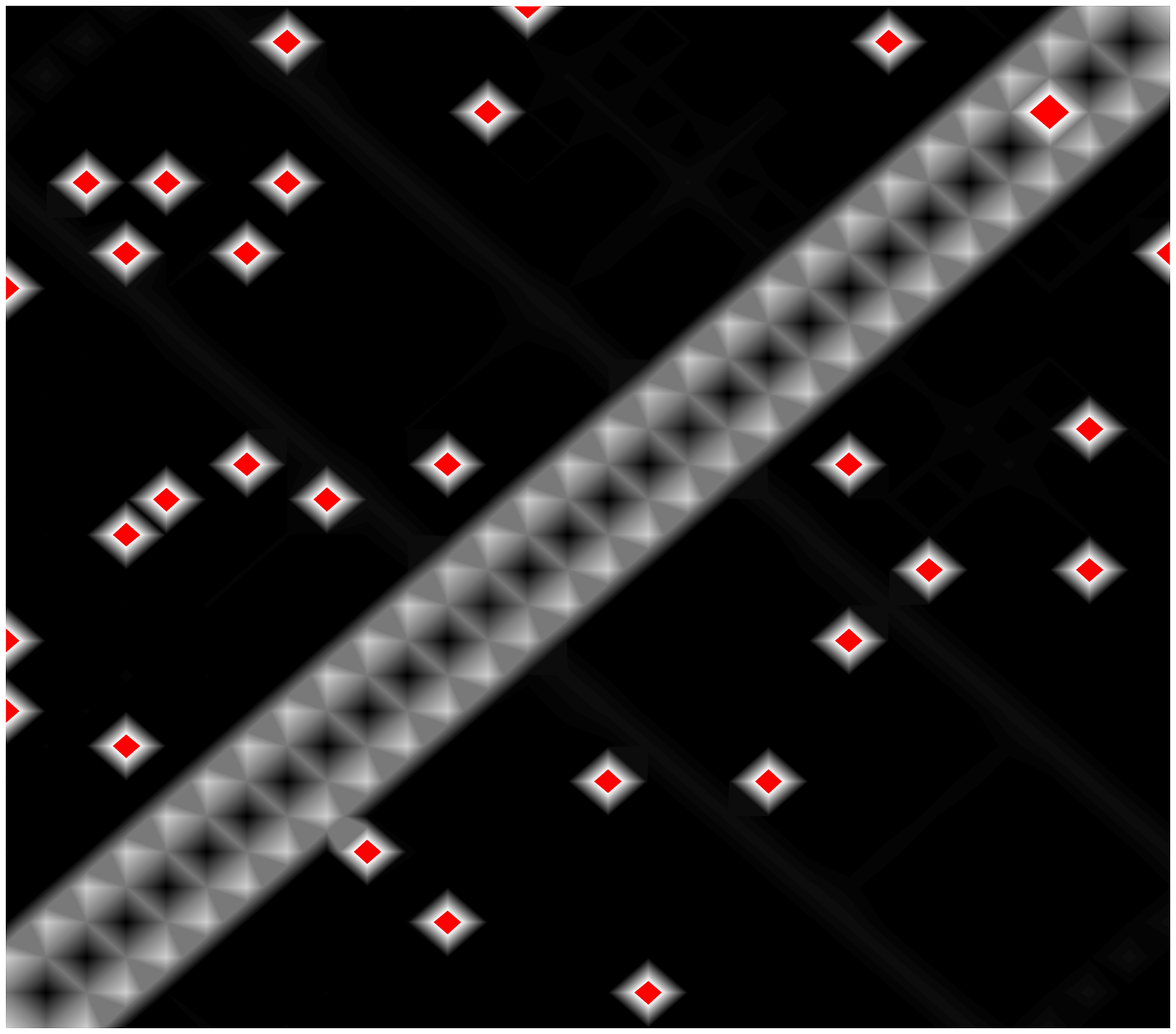} 
\includegraphics[width=4in,height=3in]{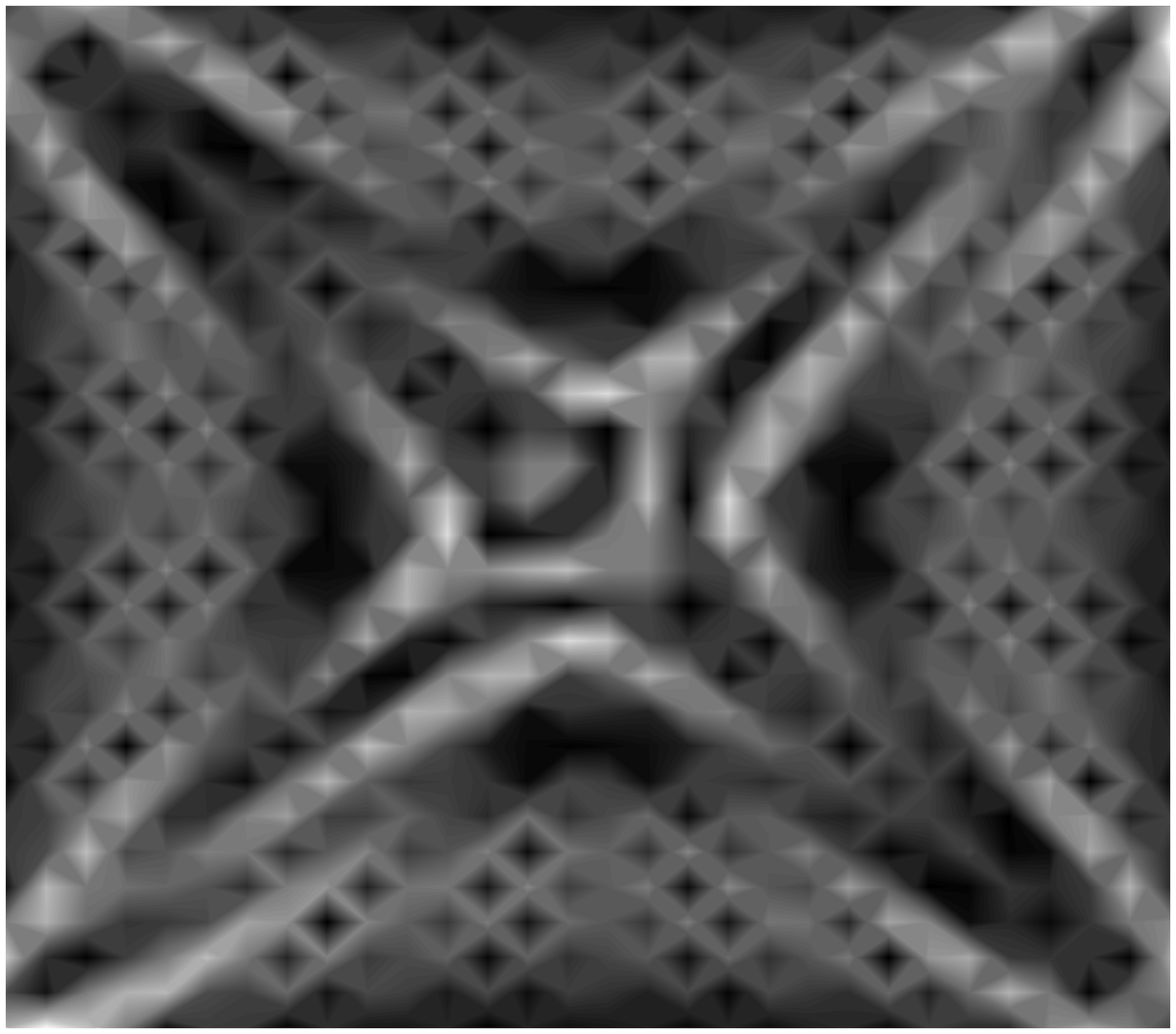} 
\caption{Probability density associated to the eigenfunction that gives  unit transmission in a $30 \times 30$ cluster connected to leads of width $W=1$
 at opposite corners of the cavity. 
Upper panel: zero magnetic flux and no vacancies. Middle panel:  zero magnetic flux and 30 vacancies 
(also shown in the figure by means of rhombi). Lower panel: same as upper panel with a magnetic flux 
$\Phi=0.1\Phi_0$ (in this case transmission is no longer 1, see text).}
\label{trajectories}
\end{figure}

\noindent {\it Classical trajectories.} At this stage we wish to identify  the nature of the state that, at zero magnetic field,  contributes to the current with exactly one conductance quantum in the case of leads
of width $W=1$ (see below). Working within the Green functions formalism 
does not allow to single out a given state as Green functions contain  information of all states at a particular energy. Then, as and it is evident that only one of the states at $E=0$ is participating in the current we should look for an alternative procedure.  We have chosen trying to identify which Hamiltonian contains the information we are looking for. In order to center the discussion let us focus on a defect free $L \times L$ cluster with leads of width 1  connected at opposite corners of the cavity.
Diagonalizing such a cluster with two negative imaginary parts added to the diagonal energies of the sites at which the leads are connected, gives $L-1$ eigenstates with
real energy at $E=0$. Displacing the leads from the corners
reduces the number of these eigenstates. None of the probability densities associated to these states seem to be related to the unit transmission found in this case. What actually happens is that, in order to build up an eigenstate which travels from one lead to the other,
one should diagonalize a Hamiltonian which includes a large portion of the leads. When this is done, the eigenstate missing from the $L$ set shows up. Now, to establish a current we have added to the diagonal energies at the
two lead ends a positive and a negative imaginary part, respectively, which is a practical way to introduce the source and sink required  to produce a current.
This procedure splits an eigenstate from the $L$ degenerate set at $E=0$ which is precisely the one that contributes to the current. The result obtained on a 
$30 \times 30$ defect free cluster connected to leads of length 100 is shown in the upper panel of Fig. \ref{trajectories}.  The figure
shows the probability density associated to the just mentioned
eigenstate. Interestingly enough it is noted that the image
is much like a classic ray. We have checked that all of
the rest of the $L-1$ eigenstates remain  localized within the cavity.
 Introducing vacancies slightly distorts this straight trajectory but still keeping the classical picture we have in mind. The results for the same cavity with 30 vacancies are shown in the middle panel of Fig. \ref{trajectories}. The system has enough degrees of freedom (see below) to build up a trajectory that eludes the blocking vacancies. Of course, as this is not always possible, the average conductance is reduced by the presence of vacancies (see  below). 
Nonetheless, the classical picture of  Fig. \ref{trajectories} offers an explanation to the odd conductance distribution of Fig. \ref{pd_6} reduced to
just two peaks at 0 and $G_0$.  However, as shown in the lower panel of Fig. \ref{trajectories}, 
a magnetic field put in evidence the quantum signature of that state, 
destroying the classical picture and, as a consequence, 
the transmission is reduced.  Thus, the conductance histogram 
(not shown here) is no longer similar to that of Fig. \ref{pd_6}.
How are these results related to the absence of the weak
localization effect shown in Fig. \ref{fi_78_6}?.
As  already observed in \cite{JB90},
the change in $G$ for small magnetic fields is negligible 
small unless a ``stopper'' is placed on the cavity
to prevent direct transmission. On semiclassical grounds, the magnitude
of the weak-localization correction was derived to be \cite{TN97}
$ G(\Phi) - G(0) \propto T_{cl}  R_{cl}$ where $T_{cl}$ and $R_{cl}$ are the
classical transmission and reflection probabilities respectively. Therefore
if  a trajectory  eludes the vacancies (as in Fig. \ref{trajectories}),  
$R_{cl} \sim 0$ and  the weak localization correction is  suppressed.
This explains  the absence of the weak localization effect.
Besides, the spreading of the probability amplitude shown in  
Fig. \ref{trajectories}, is the cause of the decrease in conductance as the 
field is switched on. 
\noindent 

{\it Discussion}.  Now the question is how can the system manage to build 
up a wavefunction that avoids scattering with vacancies. Building up a 
wavefunction localized along lines requires combinations of a sufficiently 
large number of momenta. This is actually possible due to the $L$ available 
momenta with $k_x+k_y=\pi$ that, in the defect free cavity, 
are associated to the eigenstates at $E=0$. 
By combining the wave functions of this linear space in the reciprocal 
lattice, and others with energy positive {\it and} negative very close 
to zero, it is possible to construct trayectories in real space which 
are straight lines parallel or perpendicular to that space. 
States with energies 
positive and negative participate with the same weight, 
to guarantee that the total energy is zero. 
This has been verified by projecting the wavefunctions of Fig. 4 
(upper and middle panel)
onto the eigenstates of the isolated cavity.
When  leads of width $W=1$ are attached to the cavity, there are  
leads positions for which there is a perfect matching between the 
wavefunctions of
the cavity and those of the semi-infinite leads and, thus, perfect
transmission. This occurs more rarely as the leads width increases.
For other energies, the manifold of constant energy 
correspond to an arbitrary curve in  ${\bf k}$-space. 
In addition, the energies in its the vecinity are 
not symetrically distributed. All this does not allow, in general, 
to construct one dimensional curves in real space. 
Is the chiral symmetry  of the lattice which  guarantees that this 
manifold is a straight line just at $E=0$. 
\begin{figure} 
\includegraphics[width=2.5in,height=2.2in]{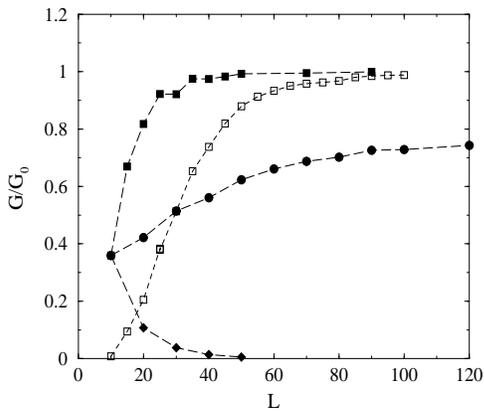} 
\caption{Average conductance  at the band center in $L \times L$ clusters 
of the square lattice with
leads of width 1 connected at opposite corners, versus 
cluster size. The results correspond to: a fixed number
of vacancies $N_v$ ($N_v=10$ filled squares, and $N_v=30$ empty squares)
$N_v=L$ (filled circles) and $N_v=0.1*L^2$ (filled diamonds).
The number of disorder realizations included in the calculations was
$10L^2$.
\label{cond_scaling}} 
\end{figure} 
In three dimensions (a cubic lattice) a similar effect shows up. 
If leads of width
$W=1$ are connected to opposite corners of a cube of side $L$ the 
conductance is exacly one quantum. It can be easily checked that
in such a cube there are $3L-2$ eigenstates with zero energy contained
in the plane $k_x+k_y+k_z=3\pi/2$. However, this only holds for $L$ odd.

\noindent {\it Scaling}.  It is interesting to investigate how the
above results scale with the system size. As shown in Fig. \ref{cond_scaling},
when the number of vacancies is constant the conductance  increases
with $L$ reaching 1 at a size which depends on the actual number of vacancies.
This shows that as far as the number of eigenstates at $E=0$ removed
by the defects remains finite, the fraction of disorder realizations 
for which the conductance is zero will vanish. 
If the number of vacancies is proportional
to the linear size of the system (in the figure we show results for
$L$ vacancies) $G$ again increases with $L$ although our
results ddo not elucidate
whether it reaches unity for large systems. Finally, for a constant
defect concentration the conductance goes exponentially to zero 
as in the case of leads of width $W=L$ \cite{CS03}.

\noindent{\it Experimental implementation}.
Is it possible to detect this effect? In recent years quantum corrals 
have been assembled 
by depositing a closed line of atoms or molecules on noble metal surfaces 
\cite{ES00,CL93,He94,ML00}.  In these sistems the local  density 
of states (DOS) reveals patterns that remind the wave functions of 
two dimensional non-interacting electrons under the corresponding 
confinement potential. 
These systems would be canditates in order to detect the effect 
here described. The transport must be done now in the plane of the 
surface, not orthogonal to it.  Also, Co atoms can be inserted into the 
quantum corral. When this atom is in the Kondo regime produces a local 
depletion in the density of states  which can simulate the effect of  
vacancies in our model, allowing to made up arbitrary distributions 
(in  number and location) of them.  Another possibility is to build up 
a dot array \cite{WL97} on a square lattice with a number of electrons that
exactly half fill one of the bands of the array. Controlling the gate 
potentialapplied to a particular dot, it can be put into or out resonance, 
simulating in this way the existence of vacancies in the array.

Partial financial support by the spanish MCYT (grant  MAT2002-04429-C03),the argentinians UBACYT (x210 and x447) and
Foundaci\'on Antorchas, and the
University of Alicante are gratefully acknowledged. We are thankful
to C. Tejedor for interesting remarks.  
  
\end{document}